\documentclass[]{aastex631}
\usepackage{amsmath}
\usepackage{natbib}

\begin{document}
\nolinenumbers
\title{Detection of a peculiar noise type in the TESS "fast" light curves}

\author[0000-0003-3754-7889]{Szil\'ard K\'alm\'an}
\affiliation{Konkoly Observatory, HUN-REN Research Centre for Astronomy and Earth Sciences, Konkoly Thege 15-17, 1121 Budapest, Hungary}
\affiliation{HUN-REN CSFK, MTA Centre of Excellence, Budapest, Hungary}
\affiliation{HUN-REN-ELTE Exoplanet Research Group, Szombathely, Szent Imre h. u. 112., H-9700, Hungary}
\affiliation{ELTE E{\"o}tv{\"o}s Lor\'and University, Doctoral School of Physics,  Budapest, Pázmány Péter sétány 1/A, H-1117, Hungary}
\author[0000-0001-6803-9698]{Szil\'ard Csizmadia}
\affiliation{German Aerospace Center (DLR), Institute for Planetary Research, Department of Extrasolar Planets and Atmospheres, 12489 Berlin, Rutherfordstrasse 2., Berlin, Germany}
\author{Andr\'as P\'al}
\author[0000-0002-0606-7930]{Gyula M. Szab\'o}
\affiliation{ELTE E{\"o}tv{\"o}s Lor\'and University, Gothard Astrophysical Observatory, Szombathely, Szent Imre h. u. 112., H-9700, Hungary}

\begin{abstract}

We present the detection of a peculiar high-frequency noise component in the 20 second cadence SAP (Simple Aperture Photometry) light curve of TESS (Transiting Exoplanets Survey Satellite). This effect (labeled as blue noise) may be attributed to the pointing instability (also known as satellite jiiter) of the satellite. We present a common technique used in the mitigation of the jitter, by decorrelating against the subpixel position of the photo-center of the point spread function of the star. We also show that a simple linear or polynomial technique may not yield satisfactory corrections, as the behavior or attitude of the noise properties may change considerably throughout the light curve.

\end{abstract}


\section{Introduction} \label{sec:intro}

The undesired signals in any given time series $\epsilon (t)$ (such as a light curve), known as noise, can be conveniently characterized by their power spectral density (PSD) $S(f)$. Following Parseval's theorem, the PSD of a signal $\epsilon (t)$ over a time period $T$ is defined as
\begin{equation}
    S(f) = \lim_{T \to \infty} \frac{1}{T} \vert \hat{\epsilon }(f) \vert^2,
\end{equation}
where $\hat{\epsilon}(f)$ is the Fourier transform of $\epsilon(t)$:
\begin{equation}
    \hat{\epsilon}(f) = \int_{- \infty}^{\infty} e^{-2\pi i f t} \epsilon(t) dt.
\end{equation}
The PSD is often characterized as 
\begin{equation}
    S(f) \propto \frac{1}{f^\gamma},
\end{equation}
where the exponent $\gamma$ is used to define different noise types, associated with colors. In audio engineering, the American National Standard T1.523-2001, Telecom Glossary 2000\footnote{\url{https://its.ntia.gov/about/resources/federal-standard-1037c}} provides the definitions of certain noise types. Three of these are adopted (somewhat loosely) into astronomy.

\textbf{White noise}. A uniform PSD over a given range of frequencies ($\gamma = 0$) is referred to as white noise or `uncorrelated noise'. This is perhaps the  most fundamental noise type underpinning the modelling of wide ranges of astronomical processes. This is primarily because a wide range of instruments rely on photon counting, yielding so-called shot noise, where the PSD is known to be independent of the frequency.

\textbf{Pink noise}. When the PSD decreases at a rate of $10$ dB/decade ($\gamma = 1$), we refer to pink noise. In the study of exoplanetary systems and especially the light curves of transiting exoplanets, this noise type is known as `$1/f$' noise or (less precisely) 'red noise' \citep{pont06} and even `correlated noise'.  In realistic observations, especially those concerning transiting exoplanets, the appearance of pink noise is unavoidable. Various phenomena, related to both the activity of the host star, as well as the observing instrument \citep[see e.g.][for a detailed list of possible sources]{csizmadia23} can cause $1/f$ noise, which then increases the difficulty of obtaining precise and accurate planetary parameters \citep{pont06, kalman23_pow}. The usage of non-parametric models for the mitigation of the effects of pink noise, such as Gaussian processes \citep[GPs; e.g.][]{Rasmussen06, Roberts12} or ones based on the wavelet-transform \citep[e.g.][]{carter09}. Both of these methods are proven to be effective based on tests conducted on synthetic data \citep{barros20, csizmadia23, kalman23_pow}. %The wavelet-based noise filtering algorithm of \cite{carter09} as implemented in the Transit and Light Curve Modeller \citep[TLCM;][]{csizmadia20, csizmadia23} is found to be effective even in the presence of severe stellar activity \citep{kalman22, bokon23, kalman24_w167, smith24}. 

\textbf{Blue noise}. If the increase in the PSD is directly proportional to the frequency ($\gamma = -1$), the noise is referred to as blue. Natural phenomena that are associated with blue noise are rare. The energy spectrum of the Cherenkov radiation \citep{Cherenkov:1934ilx} has this property in the visible spectrum. Glitches in pulsars may also be associated with the appearance of high frequency features in their PSD \citep[e.g.][]{cheng87, hu20}. \cite{Reinert16} provides a list of mathematical processes that can be used to generate (synthetic) blue nose. Detections of blue noise are generally associated with instrumental effects. Unlike in the case of pink noise, binning can act as a low-pass filtering approach, and consequently can help mitigate the effect of blue noise -- at the price of decreased temporal resolution. 
To our knowledge, there are no studies exploring the effects of blue noise on the light curve modelling of transiting exoplanets.

In this paper we present the detection of blue noise in the 20 second cadence TESS light curve, by using the data of the WASP-167 as an example, which we attribute to the pointing instability of the spacecraft. We show a commonly used approach that may help mitigate this effect, however, we also argue that it may not be adequate in every circumstance.

\section{TESS photometry}

\begin{figure}
    \centering
\includegraphics[width = \textwidth]{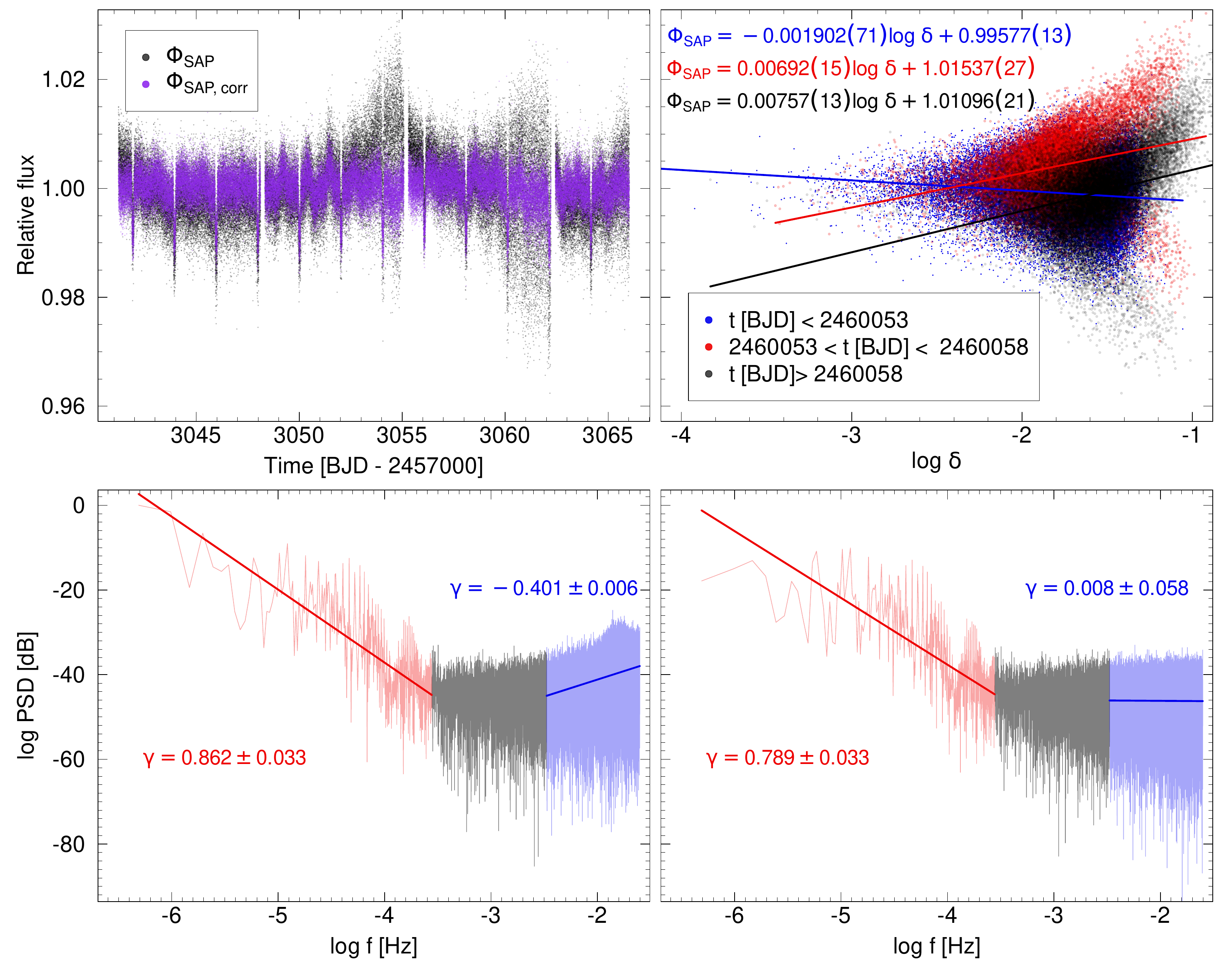}
    \caption{TESS light curve of WASP-167 (top right). The PSD from $\Phi_{\rm SAP}$ and $\Phi_{\rm SAP, corr}$ are shown in the lower left and right panels, respectively. $\Phi_{\rm SAP}$ as a function of the distance from the photo-center is shown in the upper right panel at three subsequent segments of the light curve.}
    \label{fig:fig}
\end{figure}

We obtained the TESS \citep[Transiting Exoplanets Survey Satellite;][]{ricker15} SAP (Simple Aperture Photometry) light curves of the WASP-167 \citep{temple17} system from Sector 64. Given that the blue noise features are significant in the high frequency regime, we opted to analyze data observed with the lowest available cadence, $20$ seconds. We corrected the dataset for the effects of crowding via
\begin{align}
    \Phi_{\rm SAP} (t_i) &= \frac{\Phi_{\rm SAP, raw}(t_i) - (1-c) \cdot \hat{M}(\Phi_{\rm SAP, raw})}{f}\\
    \Delta \Phi_{\rm SAP}(t_i) &=\frac{\Delta \Phi_{\rm SAP, raw}(t_i)}{f},
\end{align}
where $\Phi_{\rm SAP, raw}$ and $\Delta \Phi_{\rm SAP, raw}$ denote the raw SAP (Simple Aperture Photometry) flux and its error, $c$ is the value of the \verb|CROWDSAP| parameter (estimating the flux ratio of the target to the total flux within the aperture) and $f$ corresponds to the \verb|FLFCSAP| metric, estimating the portion of the flux of the object that is included in the aperture. The median operator is denoted by $\hat{M}$. %\textcolor{red}{reference?}

The TESS pipeline also estimates the pixel coordinates of the photo-center (of the point spread function of the star) within the aperture. Let us denote the position at every timestamp $t_i$ as $\delta_x$ and $\delta_y$. By using these values as a proxy for the telescope positioning, we apply a correction for the pointing instability. 
This effect is commonly corrected for via a bilinear fit 
\begin{equation} \label{eq:sap}
    \Phi_{\rm SAP, corr}(t_i) = \Phi_{\rm SAP}(t_i) - a \cdot \delta_x (t_i) - b \cdot \delta_y(t_i).
\end{equation}
or with similar higher-order polynomial expressions \citep{2014PASP..126..948V}. The contamination- and jitter-corrected light curves are shown on Fig. \ref{fig:fig}.

% First, we established a bilinear relationship between the crowding-corrected light curve and the ($\delta X$, $\delta Y$) values as
% \begin{equation}
%     \Phi_{\rm corr}(t_i) = a\cdot \delta_x (t_i) + b \cdot \delta_y (t_i) + c.
% \end{equation}
% The coefficients $a$, $b$, and $c$ can be estimated via a least-squares regression. The position-corrected light curve can then be given as
% \begin{equation}
%     F_{\rm corr, pos}(t_i) = F_{\rm corr}(t_i) - a \cdot \delta X (t_i) - b \cdot \delta Y(t_i).
% \end{equation}

\section{Results \& conclusion} \label{sec:results}

We compute\textcolor{red}{d} the power spectral density of $\Phi_{\rm SAP}$ and $\Phi_{\rm SAP, corr}$. These are shown on the lower left and right panels of Fig. \ref{fig:fig}, respectively. The peculiar behavior of $\Phi_{\rm SAP}$ is visible towards the high frequency end of the PSD. We divide the spectrum into three parts: the low-frequency domain (at timescales longer than 1 h, containing the stellar signals as well as the transits and phase curve effects, Fig. \ref{fig:fig} lower row, red), the high-frequency domain (with timescales shorter than 5 min, Fig. \ref{fig:fig} lower row, blue), and the medium frequency domain in between. We fit linear models on the power spectral density on the low and high frequency regimes to measure $\gamma$. We expect that at higher frequencies, only white noise (i.e. $\gamma = 0$) is detectable. We measure $\gamma = 0.862 \pm 0.033$ and $\gamma = 0.789 \pm 0.033$ at the red end of the spectrum for $\Phi_{\rm SAP}$ and $\Phi_{\rm SAP, corr}$, respectively. This is consistent with the so-called red noise, and the well-established noise filtering algorithms \citep[e.g.][]{carter09} are found to be able to handle these effects \citep[see e.g.][where this particular light curve is analyzed]{kalman24_w167}. At the high frequency end of the PSD of $\Phi_{\rm SAP}$, we measure $\gamma = -0.401 \pm 0.006$, which hints at the presence of blue noise. The wavelet-based noise filtering approach of \cite{carter09} assumes $\gamma = 1$ for the correlated noise component. While technically it could be feasible to adjust $\gamma$ as a fitting parameter, it would likely lead to degeneracies and overfitting. We find that the jitter correction can mitigate the high-frequency noise component (Fig. \ref{fig:fig}, lower right panel). Consequently, we suggest that sub-pixel sensitivity variations on the CCD may be behind the presence of blue noise in the light curves. Determining how this noise type can affect the parameter retrieval of exoplanetary systems is beyond the scope of this paper.

However, a bilinear relationship (Eq. (\ref{eq:sap}) may be inadequate in handling the effects of the pointing instability. To demonstrate this, first we compute the total distance of the photo-center $\delta = \sqrt{\delta_x^2 + \delta_y^2}$, measured from the mean pixel coordinates of $\delta_x$ and $\delta_y$. Then we show $\Phi_{SAP}$ as a function of $d$ (Fig. \ref{fig:fig}). In an ideal case, we expect this distribution to be close to a two-dimensional Gaussian. Instead, we identify at least three $\Phi_{\rm SAP}$ -- $d$ groups corresponding to different segments of the light curves (at $t < \textrm{BJD}\, 2460053$, at $\textrm{BJD}\,  2640053 < t_i < \textrm{BJD}\,  2460058$, and at $t  > \textrm{BJD}\, 2460058$). By fitting linear models in the $\Phi_{\rm SAP}$ -- $\log d$ plane, we establish significantly differing relationships in these three groups (Fig. \ref{fig:fig}, upper right panel). We find 
\begin{align}
    \Phi_{\rm SAP} &= (-0.001902 \pm 0.000071) \log \delta + 0.99577 \pm 0.00013, ~\text{when}~t<\textrm{BJD} \,2460053 \\
    \Phi_{\rm SAP} &= (0.00692 \pm 0.000015) \log \delta + 1.01537 \pm 0.00027,~ \text{when}~\textrm{BJD} \,2460053 < t < \textrm{BJD}\,2460058 \\
     \Phi_{\rm SAP} &= (0.00757 \pm 0.000013) \log \delta + 1.01096 \pm 0.00021, ~\text{when}~t>\textrm{BJD} \,2460058. 
\end{align}
We argue that the (correlated) noise level may therefore change throughout the light curve, and that the 20 s SAP light curves should be treated with special care to the telescope jitter. It is possible that using a constant amplitudes to describe the non-white noise may yield improper noise estimation. 

Beyond the example of the WASP-167 light curve presented in this paper, similar high-frequency features can be detected in the 20-second-cadence light curve of WASP-121 from sector 61, KELT-25 from sector 61, but not in the light curve of HD 1897333 from sector 81. The feature is therefore neither unique to WASP-167, nor ubiquitous for every fast light curve. Although certain physical processes occuring in the atmosphere of a planet (such as lightning) may have timescales shorter then the available exposure time \citep[e.g.][]{2019A&A...621A.113F}, which may manifest as the observed blue noise shown in Fig. \ref{fig:fig} due to the limitation of the Nyquist frequency, the peculiar noise type is unlikely to have an astrophysical origin. A deeper exploration of the root causes is beyond the scope if this work. %This effect may not be unique to the TESS satellite -- PLATO \citep[PLAnetary Transits and Oscillations of stars;][]{rauer24} may also 

%A detailed investigation of the parameter retriaval of the selected planets is beyond the scope of this paper. For that reason, we adopt the parameter sets from \cite{kalman24_w167} for WASP-167b, and \cite{bourrier20} for WASP-121b and kept these fixed throughout the light curve modelling. Instead, we study the properties of the PSD of $F_{\rm corr}$ and $F_{\rm corr, pos}$, and how the wavelet-based noise filtering affects these. Consequently, we only fitted the two parameters of the noise reduction technique, $\sigma_w$ and $\sigma_r$. 

\begin{acknowledgements}
\nolinenumbers
Project no. C1746651 has been implemented with the support provided by the Ministry of Culture and Innovation of Hungary from the National Research, Development and Innovation Fund, financed under the NVKDP-2021 funding scheme. The data are available in the Mikulski Archive for Space Telescopes \citep{https://doi.org/10.17909/t9-st5g-3177}. This project has been supported by the SNN-147362 grant  of  the  Hungarian  Research,  Development  and  Innovation  Office  (NKFIH).
\end{acknowledgements}
\bibliography{sample631}{}
\bibliographystyle{aasjournal}

%% This command is needed to show the entire author+affiliation list when
%% the collaboration and author truncation commands are used.  It has to
%% go at the end of the manuscript.
%\allauthors

%% Include this line if you are using the \added, \replaced, \deleted
%% commands to see a summary list of all changes at the end of the article.
%\listofchanges

\end{document}